\documentclass[sigconf,authorversion]{acmart}
\usepackage{acronym}
\usepackage{subfigure}

\acrodef{MIUR}{Ministry of University and Research}
\acrodef{SD}{Scientific Discipline}
\acrodef{SA}{Scientific Area}

\acrodef{CHE}{Chemistry}
\acrodef{MCS}{Mathematics and Computer Sciences}
\acrodef{PHY}{Physics}
\acrodef{EAS}{Earth Sciences}
\acrodef{BIO}{Biology}
\acrodef{MED}{Medical Sciences}
\acrodef{AVM}{Agricultural Sciences and Veterinary Medicine}
\acrodef{CEA}{Civil Engineering and Architecture}
\acrodef{IIE}{Industrial and Information Engineering}
\acrodef{APL}{Antiquities, Philology, Literary Studies, Art History}
\acrodef{HPP}{History, Philosophy, Pedagogy and Psychology}
\acrodef{LAW}{Law}
\acrodef{ECS}{Economics and Statistics}
\acrodef{PSS}{Political and Social Sciences}

\newenvironment{fivenum}{
  \begin{tabular}{lllll}
    \toprule
    {\em Min.} & {\em 1st Qu.} & {\em Median} & {\em 3rd Qu.} & {\em Max.} \\   
}{%
  \bottomrule
  \end{tabular}
}

\setcopyright{acmcopyright}
\begin{document}

\copyrightyear{2019}
\acmYear{2019}
\acmConference[ECSA]{European Conference on Software Architecture}{September 
9--13, 2019}{Paris, France}
\acmBooktitle{European Conference on Software Architecture (ECSA), September 
9--13, 2019, Paris, France}
\acmPrice{15.00}
\acmDOI{10.1145/3344948.3344966}
\acmISBN{978-1-4503-7142-1/19/09}

\title{Gender Balance in Computer Science and Engineering in Italian Universities}

\author{Moreno Marzolla}
	\affiliation{
 	\institution{University of Bologna}
	\city{Bologna} 
 	\state{Italy} 
  	\postcode{40126}}
	\email{moreno.marzolla@unibo.it}

\author{Raffaela Mirandola}
  	\affiliation{
 	\institution{Politecnico di Milano}
	\city{Milan} 
 	\state{Italy} 
  	\postcode{20133}}
  	\email{raffaela.mirandola@polimi.it}

\begin{abstract}
Multiple studies have shown that gender balance in the fields of
Science, Technology, Engineering and Maths -- and in particular in ICT
-- is still far to be achieved. Several initiatives have been recently
taken to increase the women participation, but it is difficult, at
present, to evaluate their impact and their potential of changing the
situation.  This paper contributes to the discussion by presenting a
descriptive analysis of the gender balance in Computer Science and
Computer Engineering in Italian Universities.
\end{abstract}

%
%



\maketitle

\section{Introduction}\label{sec:introduction}

In recent years, gender diversity and equality in STEM (Science,
Technology, Engineering and Mathematics) in general, and in Computer
Science and Engineering in particular, has been the subject of several
studies and
events~\cite{ge18,ge19,WSA-ECSA,McKinsey,paradox,education} that
emphasize how diverse skills could be beneficial in dealing with the
raising complexity of today's systems. This interest is motivated by
the observation that diversity enhances creativity and multi-cultural
experience leads to better outcomes~\cite{cra-w}. The results
collected so far highlight how the percentage of women working in ICT
is quite low (less than 10\% in some areas); the initiatives taken as
yet to promote the women participation in~ICT are quite at their
infancy and so it is difficult to evaluate their impact and their
potential of changing the situation.  Such initiatives include events
pointing out the historic female role models like Ada
Lovelace~\cite{festival} and Margaret Hamilton~\cite{icse2018}, awards
and fellowships schemas as the EU Marie Curie
program~\cite{mariecurie} or the Google Program Women
Techmakers~\cite{google}, to cite a few. However, some of these
measure -- if dedicated exclusively to women -- can be perceived
negatively, creating the impression that they are easier to obtain,
based on gender and not on merit.

Understanding and addressing the gender gap is a complex issue with
several facets that require the involvement not only of~ICT people
but also sociologists, psychologists and politicians. In this paper we
contribute to the discussion by analyzing the gender balance in
Computer Science and Computer Engineering in Italian universities.
Our analysis will be mostly descriptive: investigating the root causes
of the observed situation and assessing existing and possible
countermeasures is the topic of ongoing research.

The work has been inspired by a recently appeared paper, \emph{The
  gender equality paradox in STEM education}~\cite{paradox}, where the
authors found that countries with high levels of gender equality have
some of the largest~STEM gaps in secondary and tertiary education.
According to this study, the number of women in~STEM in Italy is close
to 35\%; Italy is placed around the average point with respect to the
global gender gap index, and has comparatively more women in~STEM than
other countries with the same index. We found this fact quite
intriguing, since the perception in our work environment was quite
different. To have a clear picture of the gender balance situation in
Italian universities, we retrieved and analyzed publicly available
data about the professors/researchers and of students.

The paper is organized as follows. Section~\ref{sec:related-work}
summarizes some related works dealing with the analysis and
understanding of gender issues in
STEM. Section~\ref{sec:data-and-methods} describes the dataset that
has been collected, and the analysis methods employed in the present
paper; Section~\ref{sec:analysis} discusses the results of the
analysis. Finally, conclusions and future research directions are
briefly summarized in Section~\ref{sec:conclusions}.

\section{Related Work}\label{sec:related-work}

To begin with, it is important to observe that women never left the
computing field but were pushed out: the historical perspective
illustrated in~\cite{cacm18} contains several actual examples, from
the women in the~ENIAC team whose contribution was never publicized,
to more recent cases of plain academic
hostility~\cite{Wynn:2018,McKinley:2018}. The influence of negative
media images about women in computing is described in~\cite{cacm19},
where the authors highlight how for gender-clich\`ed images
\emph{"ordinary women" are perceived as less than capable, while
  ordinary men can and do fully engage and participate in the
  computing disciplines}.

The already mentioned paper~\cite{paradox} shows that more
gender-equal societies have fewer women taking~STEM degrees, while the
percentage of women STEM graduates is higher for countries that have
more gender inequality. For example, countries like Tunisia, Albania
and Turkey with a low global gender equality index have between 35\%
and 40\% of women among STEM graduates, while in Finland and Norway,
traditionally with a high gender equality index, the percentage of
women among STEM graduates drops to 20\%.  The paper raised several
discussions about its validity, some of which concerned the usage of
the gender equality index or of the data
analyzed\footnote{\url{https://blog.raulza.me/category/gender/},
  accessed on 2019-05-27}.

In the proceedings of Gender Equality workshops held at~ICSE in May
2018~\cite{ge18} and May 2019~\cite{ge19} the readers can find the
details of several initiatives devoted to the promotion of gender
equality. In the following we briefly describe some of these
contributions.

A socio-historical study trying to explain the decreasing percentage
of women in ICT is presented in~\cite{socio19}. Using a sociological
closure theory to explain the gendering of computing, the author
focuses on the history of enrollment booms in computer science. The
results suggest that the percentage of women is affected by specific
choices made in the admissions policies and by the presentation of the
career using male-oriented images.

In~\cite{OSS19} the author studies the presence of women in~$355$ Open
Source Software (OSS) Communities. Specifically, she focuses on the
impact of ``women only spaces'' as a way to increase the presence of
women in~OSS. Indeed, research has shown that women perform better
when they can build connections and mentoring networks with other
women.

In~\cite{UK2018} the authors present the results of a survey whose aim
is to identify significant factors affecting the gender distribution
in STEM in the~UK. The paper presents statistics on the perception of
the importance of diversity for a successful organization. Then, the
attitude of women and men with respect to diversity, equality and
mentoring is analyzed; finally, the authors identify areas of
improvements and provide practical suggestions based on the answers
received from the survey.

The situation in the~UK is the topic of another
work~\cite{cardiff2018}. Women represent~$17\%$ of the workforce in
UK's ICT and~$14\%$ of the students in undergraduate Computer Science
programmes. The authors illustrate the measures implemented by the
School of Computer Science and Informatics of Cardiff University to
foster a more balanced situation by raising awareness about gender
balance and contrasting stereotypes and unconscious biases.

In~\cite{Chile2018} the authors describe the Gender Equality
Admissions Program, active at the University of Chile since~2013 to
increase the participation of women in STEM disciplines. The aim of
the program is to contrast the strong cultural stereotype existing in
Chile according to which STEM disciplines are ``for men''. Since the
the program has been in place, the number of women accepted into
engineering and science program has grown from~$19\%$ to more
than~$32\%$.

\section{Data and Methods}\label{sec:data-and-methods}

In this section we provide some key facts about the structure of the
Italian university system. There are two types of university
professors, full and associate, both of which are permanent
positions. The lowest rank of tenured assistant professor has been
replaced since~2010 by two temporary positions: Type~A and Type~B
researcher. Type~A positions last for three years and can be renewed
once; Type~B positions last for three years and can lead to a
(tenured) associate professorship. Each university professor and
researcher in Italy is bound to a specific field of study,
called~\acf{SD}. Each~\ac{SD} is assigned an alphanumeric code of the
form $AA/MN$, where~$AA$ is a numeric value from~$00$ to~$14$ denoting
the~\ac{SA}, $M$ is a letter ($A, B, C, \ldots$) representing the
\emph{macro sector}, and~$N$ is a single digit identifying a
specific~\ac{SD} within the macro sector.

\begin{table}[t]
  \caption{The~$14$ scientific areas used in the Italian university
    system. For each area we show the numeric ID, a three-letter code
    used in this paper, the name and the number of~\acp{SD} it
    contains.}\label{tab:area}

  \centering%
  \begin{small}%
    \begin{tabular}{llp{.65\columnwidth}l}
      \toprule
      {\em ID} & {\em Acron.} & {\em Area Name} & {\em \# SD} \\
      \midrule
      01 & MCS & \acl{MCS} &   7 \\
      02 & PHY & \acl{PHY} &   6 \\
      03 & CHE & \acl{CHE} &   8 \\
      04 & EAS & \acl{EAS} &   4 \\
      05 & BIO & \acl{BIO} &  13 \\
      06 & MED & \acl{MED} &  26 \\
      07 & AVM & \acl{AVM} &  14 \\
      08 & CEA & \acl{CEA} &  12 \\
      09 & IIE & \acl{IIE} &  20 \\
      10 & APL & \acl{APL} &  19 \\
      11 & HPP & \acl{HPP} &  17 \\
      12 & LAW & \acl{LAW} &  16 \\ 
      13 & ECS & \acl{ECS} &  15 \\
      14 & PSS & \acl{PSS} &   7 \\
      \midrule 
         & Total &         & 184 \\
      \bottomrule
    \end{tabular}%
  \end{small}
\end{table}

Table~\ref{tab:area} lists the~$14$ \acp{SA} and the number
of~\acp{SD} within each area. There are~$184$ \acp{SD}, whose aims and
scope are described in~\cite{dm-set-concorsuali}; an English
translation of the names of the~\acp{SD} are provided in
Appendix~\ref{app:list-sd}. The main focus of this paper will be on
Computer Science and Engineering, that correspond to the~\acp{SD}
$01/B1$ and $09/H1$, respectively.  $01/B1$ refers to Area~$01$
(Mathematics and Computer Science), macro sector~$B$ (Informatics)
and~\ac{SD}~$1$ (Informatics); $09/H1$ refers to area~$09$ (Industrial
and Information Engineering), macro sector~$H$ (Computer Engineering)
and~\ac{SD}~$1$ (Information Processing Systems).

The~\ac{MIUR} provides the list of professors and researchers employed
in public and private state-recognized universities at the end of any
chosen year
since~2000\footnote{\url{http://cercauniversita.cineca.it/php5/docenti/cerca.php},
  Accessed on 2019-04-18}. The list includes all tenured and
tenure-track personnel (full/associate professors, tenured
researchers, tenure-track \emph{type B} researchers), and non
tenure-track \emph{type A} researchers. However, data older than 2008
is severely incomplete and has been ignored in our analysis.

Aggregate statistics on students enrolled in state-recognized
universities on each academic year are available through the Student's
National Database\footnote{\url{http://anagrafe.miur.it/index.php},
  Accessed on 2019-04-18}. The data refers to the academic years
ranging from~2003/2004 to~2017/2018, and include the number of male
and female students enrolled in bachelor and master degrees, broken
down by university and field of study. 

\section{Analysis}\label{sec:analysis}

\begin{figure*}
  \centering\includegraphics[width=.87\textwidth]{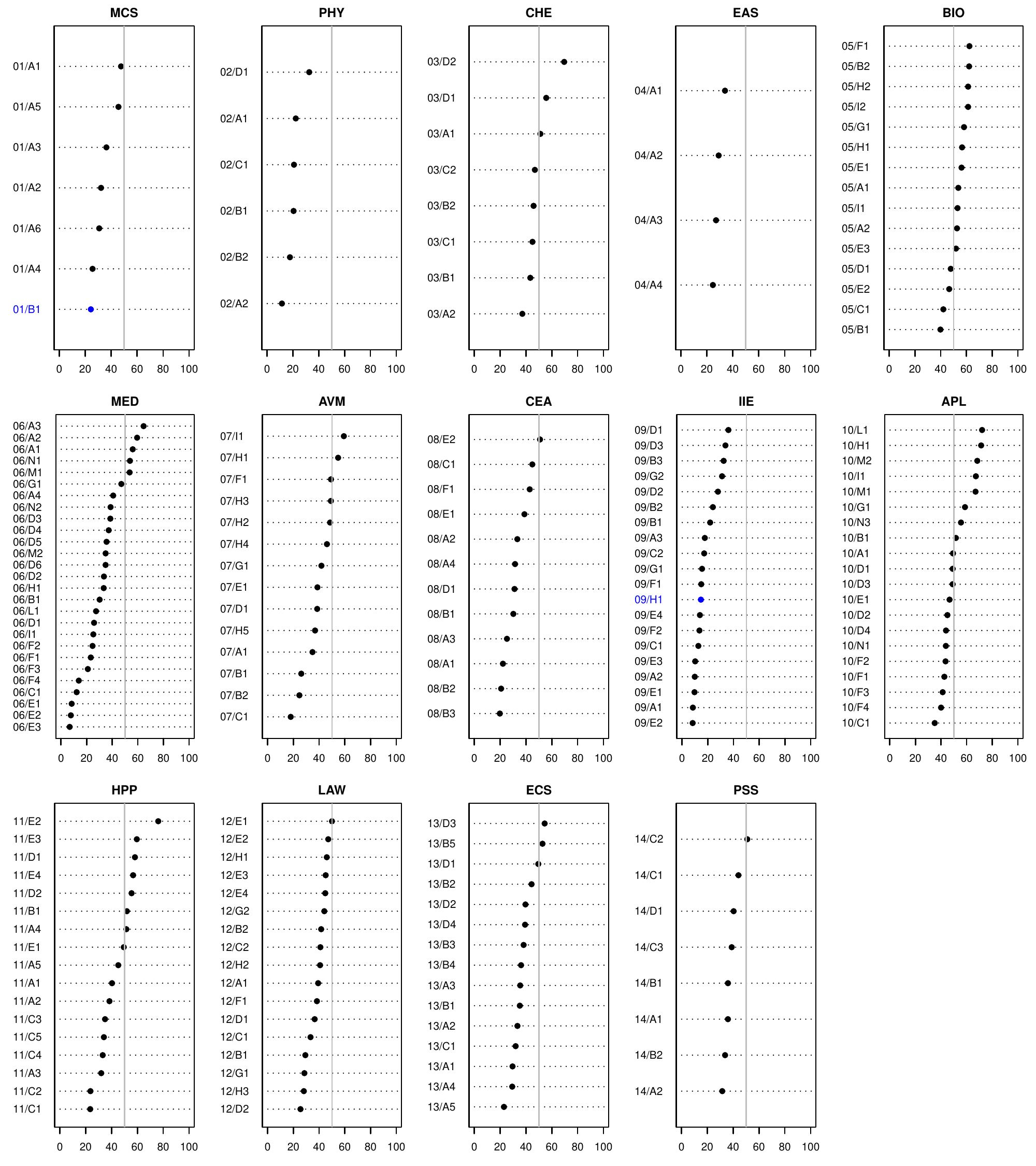}  
  \caption{Percentages of female professors and researchers in
    2018. The 01/B1--\emph{Computer Science} and
    09/H1--\emph{Information Processing Systems} \ac{SD} are shown in
    blue. Grey vertical lines are placed at the $50\%$
    mark.}\label{fig:perc-female}
\end{figure*}

For the sake of completeness, before focusing on Computer Science and
Engineering we illustrate the whole Italian situation.
Figure~\ref{fig:perc-female} shows the distribution of the percentage
of women among Italian university professors and researchers in~2018,
for each \ac{SD}. We observe that women are under-represented in many
research areas: in particular, all~\acp{SD} in the~\ac{PHY}
and~\ac{EAS} areas, and most of those in~\ac{IIE}, show a percentage
of women below~$35\%$. On the other hand, \ac{BIO} and \ac{APL} show
higher percentages of female professors and researchers. There are,
however, huge variations even within the same area. \ac{MED} is a
prominent example: the percentage of women ranges from a minimum of
$6.78\%$ in 06/E3--\emph{Neurosurgery and maxillofacial surgery} to a
maximum of $64.46\%$ in 06/A3--\emph{Microbiology and clinical
  microbiology}. It is important to remark that there is a large
variation in the number of professors in different fields of study
(see below) that might amplify the differences.

We can get a useful summary of the distribution of the percentage of
female professors across all fields of study by looking at Tukey's
five number summary~\cite{tukey1977exploratory} (minimum, first
quartile, sample median, third quartile, maximum):\medskip

\begin{center}
\begin{fivenum}
  $6.78$ & $28.25$ & $38.44$ & $48.91$ & $76.14$ \\
\end{fivenum}
\end{center}\medskip

The numbers show that:
\begin{itemize}
\item The minimum percentage of women is $6.78\%$ (in
  06/E3--\emph{Neurosurgery and maxillofacial surgery});
\item A quarter of the~\acp{SD} ($46$ out of~$184$) have a percentage
  of women $\leq 28.25\%$;
\item Half of the~\acp{SD} ($92$ out of~$184$) have a percentage of
  women $\leq 48.44\%$;
\item Three quarters of the~\acp{SD} ($138$ out of~$184$) have a
  percentage of women $\leq 48.91\%$;
\item The maximum percentage of women is $76.16\%$ (in
  11/E3--\emph{Developmental and educational psychology}).
\end{itemize}

The number of professors varies greatly across disciplines by almost
an order of magnitude. The five-number summary of the total number of
professors and researchers for each~\ac{SD} is the following:\medskip

\begin{center}
\begin{fivenum}
  $74.0$ & $174.2$ & $249.0$ & $352.5$ & $917.0$ \\
\end{fivenum}
\end{center}\medskip

The~\ac{SD} with the lowest number of professors is
13/A4--\emph{Applied Economics} ($74$ professors and researchers, $17$
of which female). Half of the~\acp{SD} have less than~$249$
professors. The~\ac{SD} with the highest number of professors is
01/B1--\emph{Computer Science} ($917$ professors, $223$ of which
female). 09/H1--\emph{Information Engineering} is the fifth largest
group with~$778$ professors, $114$ of which female.

Both the \emph{Computer Science} and \emph{Information Processing
  Systems} \ac{SD}, highlighted in blue in
figure~\ref{fig:perc-female}, score badly with respect to the
percentage of women. Computer Science has the lowest percentage
($24.32\%$) in the~\ac{MCS} research area. This value drops to
$14.65\%$ for the Information Processing Systems~\ac{SD}, that however
is not the lowest value in the~\ac{IIE} area, the lowest being
09/E2--\emph{Electrical energy engineering} ($8.26\%$ of women).

\begin{table}
  \caption{Number of male and female professors and researchers for each area}\label{tab:prof-area}
  \begin{tabular}{llrrrr}
    \toprule
    & \textbf{Area} & \textbf{F} & \textbf{M} & \textbf{Total} & \textbf{\%F} \\
    \midrule    
    01 & MCS &   $970$ &  $2073$ &  $3043$ & $31.88\%$ \\
    02 & PHY &   $476$ &  $1775$ &  $2251$ & $21.15\%$ \\
    03 & CHE &  $1353$ &  $1447$ &  $2800$ & $48.32\%$ \\
    04 & EAS &   $305$ &   $730$ &  $1035$ & $29.47\%$ \\
    05 & BIO &  $2526$ &  $2168$ &  $4694$ & $53.81\%$ \\
    06 & MED &  $3024$ &  $5922$ &  $8946$ & $33.80\%$ \\
    07 & AVM &  $1182$ &  $1822$ &  $3004$ & $39.35\%$ \\
    08 & CEA &  $1154$ &  $2302$ &  $3456$ & $33.39\%$ \\
    09 & IIE &   $980$ &  $4570$ &  $5550$ & $17.66\%$ \\
    10 & APL &  $2523$ &  $2144$ &  $4667$ & $54.06\%$ \\
    11 & HPP &  $2035$ &  $2330$ &  $4365$ & $46.62\%$ \\
    12 & LAW &  $1762$ &  $2882$ &  $4644$ & $37.94\%$ \\
    13 & ECS &  $1841$ &  $3036$ &  $4877$ & $37.75\%$ \\
    14 & PSS &   $683$ &  $1001$ &  $1684$ & $40.56\%$ \\
    \midrule
       &     & $20814$ & $34202$ & $55016$ & $37.83\%$ \\
    \bottomrule
  \end{tabular}
\end{table}

Table~\ref{tab:prof-area} shows the total number of male and female
professors on each scientific area. The three areas with the lowest
percentage of women are \acl{IIE} ($17.66\%$), \acl{PHY} ($21.15\%$)
and \acl{EAS} ($29.47\%$). On the other hand, the three areas with the
highest percentage of women are \acl{APL} ($54.06\%$), \acl{CHE}
($48.32\%$) and \acl{HPP} ($46.62\%$). It is therefore quite evident
that, in general, women are under-represented in physics and
engineering.

\begin{figure}
  \centering\includegraphics[width=.9\columnwidth]{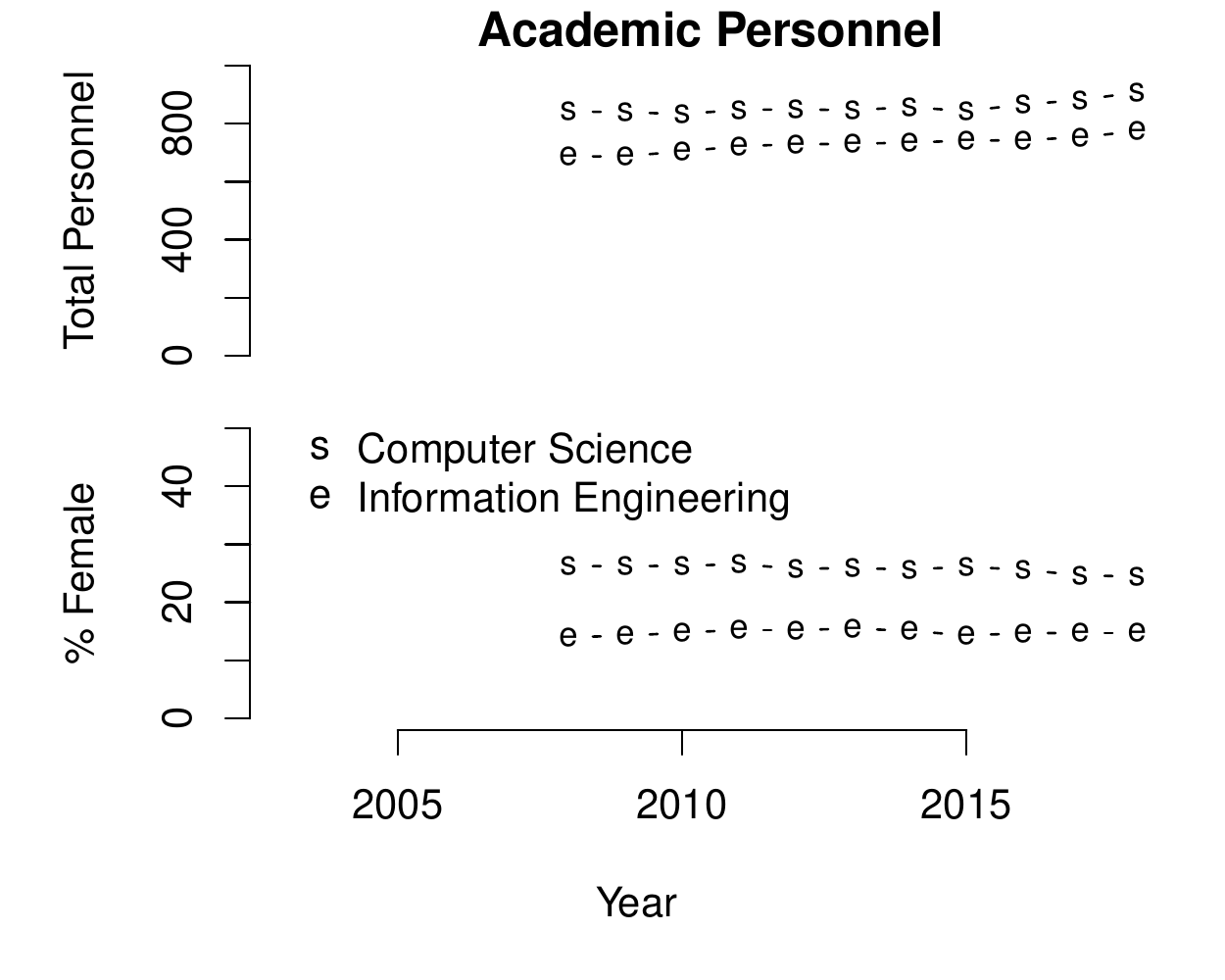}
  \caption{Academic personnel of the Computer Science and Information
    Engineering~\acp{SD} and percentages of women}\label{fig:serie-storica-docenti}
\end{figure}

Focusing our attention on Computer Science and Engineering,
Figure~\ref{fig:serie-storica-docenti} shows the historical trend of
the total number of professors in the 01/B1--\emph{Computer Science}
and 09/H1--\emph{Information Engineering}~\acp{SD}, and the percentage
of women. Computer Science has, on average, about twice the fraction
of female professors and researchers than Information
Engineering. Although the total number of professors in both
disciplines is more or less steadily increasing during the time period
under consideration (2008--today), the percentage of women remains
basically unchanged.

\begin{figure}
  \centering%
  \subfigure[\label{fig:serie-storica-bachelor}]{\includegraphics[width=.9\columnwidth]{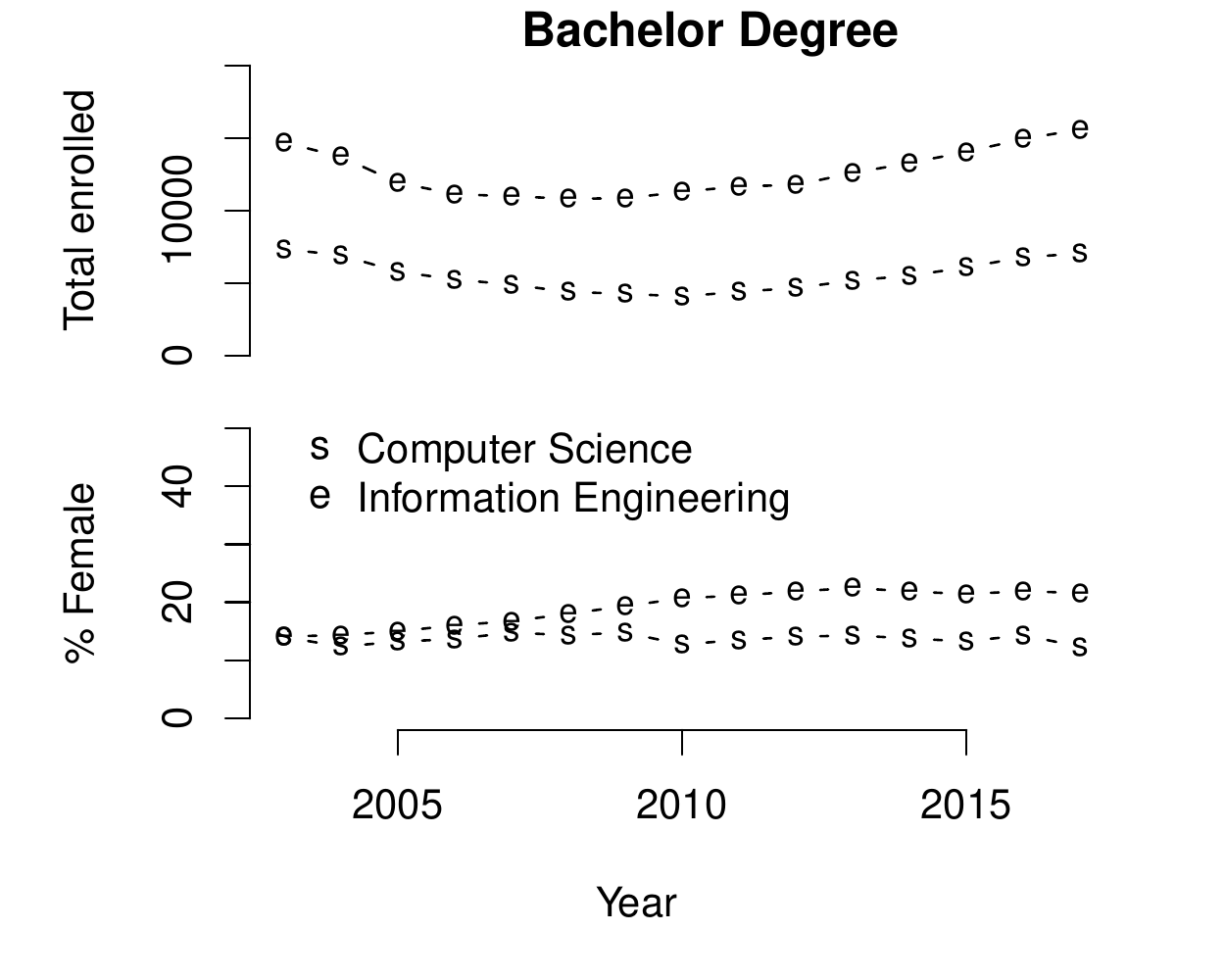}}
  \subfigure[\label{fig:serie-storica-master}]{\includegraphics[width=.9\columnwidth]{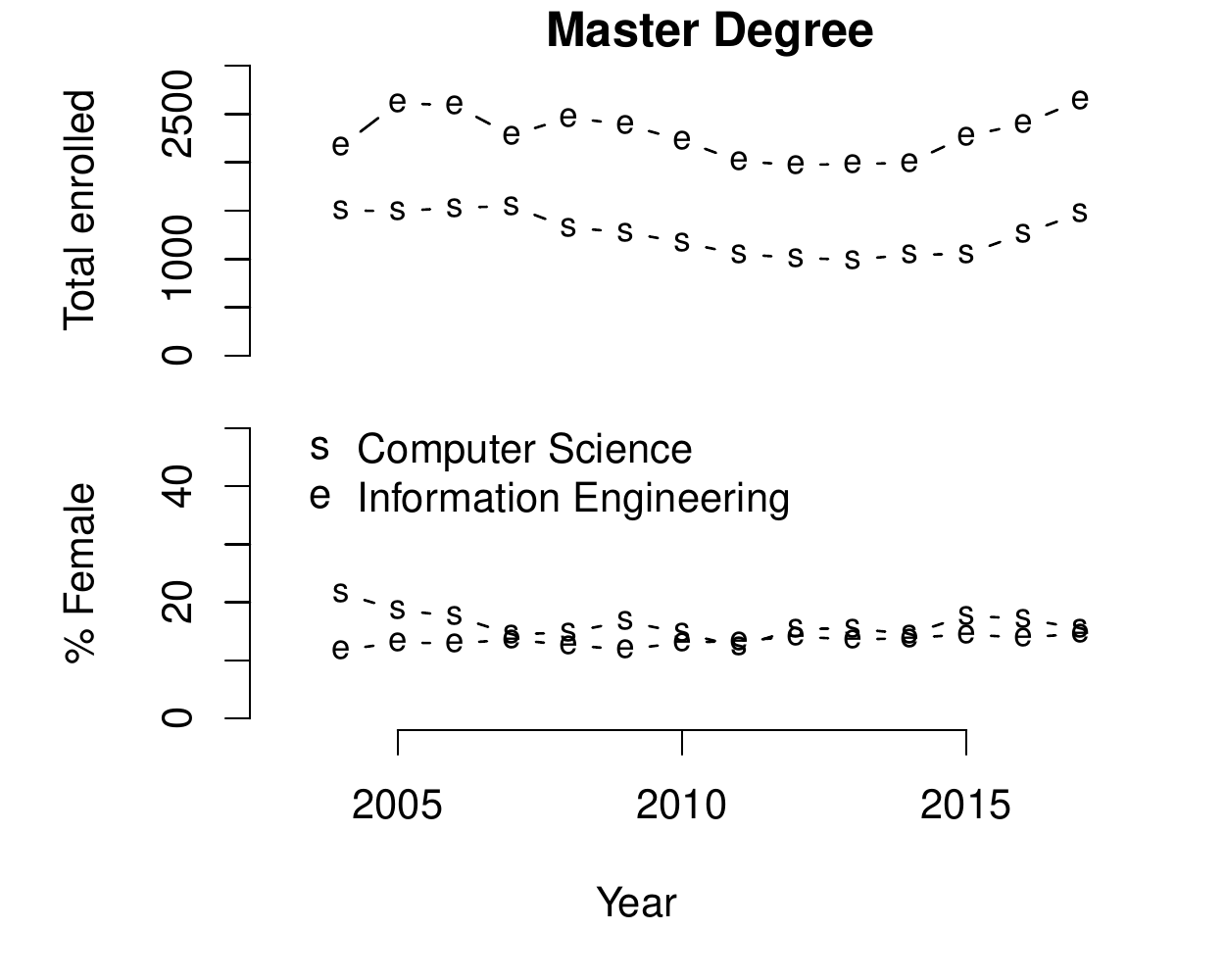}}
  \caption{Number of students enrolled in Bachelor and Master degrees
    in Computer Science and Information Engineering and percentages of
    women; note that vertical scales differ.}\label{fig:serie-storica-studenti}
\end{figure}

Figure~\ref{fig:serie-storica-studenti} shows the number of newly
enrolled students in Bachelor and Master degrees in Computer Science
and Information Engineering during the period 2003--2017. The number
of students of both degrees (Figure~\ref{fig:serie-storica-bachelor})
dropped to a minimum around the year 2009; specifically, the minimum
number of enrollments in Computer Science was~$4096$ students
in~2010, $521$ of which female, while for Information Engineering the
minimum number of enrollments was in~2008 with~$10830$ students,
$1952$ of which female. After that, yearly enrollments started to grow
again up to the most recent values of~$7096$ for Computer Science
and~$15596$ for Information Engineering, both in~2017. Interestingly,
while the percentage of female students in Computer Science is stable,
the percentage of female students in Information Engineering has grown
during the period 2005--2010, raising from about~$14\%$ to
about~$21\%$, and keeping stable since then. The last available data
point shows that, in 2017, $21.61\%$ of students in Information
Engineering are women, compared with just $12.34\%$ for Computer
Science. We currently have no plausible explanation for this
phenomenon, also considering that the percentage of female professors
of Computer Engineering is actually \emph{lower} than Computer
Science.

Figure~\ref{fig:serie-storica-master} shows the gender composition
within the newly enrolled students in Master degrees in Computer
Science and Engineering. We observe that the yearly enrollments reach
a minimum in~2013 for Computer Science ($992$ students, $154$ of which
female) and in~2012 for Information Engineering ($1974$ students,
$279$ of which female). Essentially, the negative peak of enrollments
in Bachelor degrees impacted those to the Master degrees after a few
years, presumably the time taken to students to complete the lower
degree. Looking at the percentage of women among newly enrolled
students, we observe that the fraction of female students in
Information Engineering increased slowly from~$11.86\%$ in 2004
to~$14.63\%$ in 2017. On the other hand, the fraction of female
students in Computer Science dropped a few points from~$21.61\%$
in~2004 to to about $14--15\%$ from~2007 onward, up to the most
recent value~$15.50\%$ in~2017. Interestingly, the prominent gap
between the percentages of female students in Computer Science and
Computer Engineering observed in
Figure~\ref{fig:serie-storica-bachelor} is absent on the data for
Master students. 

\section{Conclusions}\label{sec:conclusions}

In this paper we presented an initial investigation of the gender
(un-)balance situation in Italian universities, with particular
emphasis on Computer Science and Computer Engineering. We analyzed
publicly available data provided by the Italian~\acl{MIUR} about the
population of professors and students. The data shows that women are
in general under-represented among university professors, in
particular in STEM disciplines. However, the analysis of
individual~\acp{SD} shows a considerable variability within the same
area: for example, within the Medical Sciences area the percentage of
women ranges from~$6.78\%$ to~$64.46\%$, depending on the specific
field of study. Women in Computer Science and Computer Engineering are
under-represented, the situation being worse in Computer Engineering,
where in~2018 only~$14.65\%$ of professors and researchers are women,
compared with~$24.32\%$ for Computer Science.  Interestingly, the
situation is the opposite among students, where there is a lower
percentage of female students in Computer Science than Information
Engineering.

This work opens several investigation lines. We are currently
attempting to investigate the root causes of the severe gender
imbalance that is observed in Computer Science and
Engineering. Additionally, we are assessing the effectiveness of the
initiatives that have been put in place to attract more women towards
STEM disciplines. Finally, we are considering to extend the study to
different countries trying to understand possible societal influences
in the percentage of women in Computer science and Engineering.

\begin{acks}
Raffaela Mirandola has been partially supported by the
\grantsponsor{KKS}{Swedish KK-Stiftelsens}{http://www.kks.se/} project
No.~\grantnum{KKS}{20170232}.
\end{acks}

\appendix

\section{List of Scientific Disciplines}\label{app:list-sd}

The list below enumerates all scientific areas, macro-sectors and
scientific disciplines used in the Italian university system. We use
the English translation by the Italian National University
Council\footnote{\url{https://www.cun.it/documentazione/academic-fields-and-disciplines-list/},
  accessed on 2019-05-29}, since the officianl denominations are in
Italian..\medskip

\begin{description}\item[01] Mathematics and computer sciences
\begin{description}
\item[01/A] Mathematics
\begin{description}
\item[01/A1] Mathematical logic, mathematics education and history of mathematics
\item[01/A2] Geometry and algebra
\item[01/A3] Mathematical analysis, probability and statistics
\item[01/A4] Mathematical physics
\item[01/A5] Numerical analysis
\item[01/A6] Operational research
\end{description}
\item[01/B] Informatics
\begin{description}
\item[01/B1] Informatics
\end{description}
\end{description}
\item[02] Physics
\begin{description}
\item[02/A] Physics of fundamental interactions
\begin{description}
\item[02/A1] Experimental physics of fundamental interactions
\item[02/A2] Theoretical physics of fundamental interactions
\end{description}
\item[02/B] Physics of matter
\begin{description}
\item[02/B1] Experimental physics of matter
\item[02/B2] Theoretical physics of matter
\item[02/B3] Applied physics
\end{description}
\item[02/C] Astronomy, astrophysics, Earth and planetary physics
\begin{description}
\item[02/C1] Astronomy, astrophysics, Earth and planetary physics
\end{description}
\end{description}
\item[03] Chemistry
\begin{description}
\item[03/A] Analytical and physical chemistry
\begin{description}
\item[03/A1] Analytical chemistry
\item[03/A2] Models and methods for chemistry
\end{description}
\item[03/B] Inorganic chemistry and applied technologies
\begin{description}
\item[03/B1] Principles of chemistry and inorganic systems
\item[03/B2] Chemical basis of technology applications
\end{description}
\item[03/C] Organic, industrial and applied chemistry
\begin{description}
\item[03/C1] Organic chemistry
\item[03/C2] Industrial and applied chemistry
\end{description}
\item[03/D] Medicinal and food chemistry and applied technologies
\begin{description}
\item[03/D1] Medicinal, toxicological and nutritional chemistry and applied technologies
\item[03/D2] Drug technology, socioeconomics and regulations
\end{description}
\end{description}
\item[04] Earth sciences
\begin{description}
\item[04/A] Earth sciences
\begin{description}
\item[04/A1] Geochemistry, mineralogy, petrology, volcanology, Earth resources and applications
\item[04/A2] Structural geology, stratigraphy, sedimentology and paleontology
\item[04/A3] Applied geology, physical geography and geomorphology
\item[04/A4] Geophysics
\end{description}
\end{description}
\item[05] Biology
\begin{description}
\item[05/A] Plant biology
\begin{description}
\item[05/A1] Botany
\item[05/A2] Plant physiology
\end{description}
\item[05/B] Animal biology and anthropology
\begin{description}
\item[05/B1] Zoology and anthropology
\item[05/B2] Comparative anatomy and cytology
\end{description}
\item[05/C] Ecology
\begin{description}
\item[05/C1] Ecology
\end{description}
\item[05/D] Physiology
\begin{description}
\item[05/D1] Physiology
\end{description}
\item[05/E] Experimental and clinical biochemistry and molecular biology
\begin{description}
\item[05/E1] General biochemistry and clinical biochemistry
\item[05/E2] Molecular biology
\end{description}
\item[05/F] Experimental biology
\begin{description}
\item[05/F1] Experimental biology
\end{description}
\item[05/G] Experimental and clinical pharmacology
\begin{description}
\item[05/G1] Pharmacology, clinical pharmacology and pharmacognosy
\end{description}
\item[05/H] Human anatomy and histology
\begin{description}
\item[05/H1] Human anatomy
\item[05/H2] Histology
\end{description}
\item[05/I] Genetics and microbiology
\begin{description}
\item[05/I1] Genetics and microbiology
\end{description}
\end{description}
\item[06] Medicine
\begin{description}
\item[06/A] Pathology and laboratory medicine
\begin{description}
\item[06/A1] Medical genetics
\item[06/A2] Experimental medicine, pathophysiology and clinical pathology
\item[06/A3] Microbiology and clinical microbiology
\item[06/A4] Pathology
\end{description}
\item[06/B] General clinical medicine
\begin{description}
\item[06/B1] Internal medicine
\end{description}
\item[06/C] General clinical surgery
\begin{description}
\item[06/C1] General surgery
\end{description}
\item[06/D] Specialized clinical medicine
\begin{description}
\item[06/D1] Cardiovascular and respiratory diseases
\item[06/D2] Endocrinology, nephrology, food and wellness sciences
\item[06/D3] Blood diseases, oncology and rheumatology
\item[06/D4] Skin, contagious and gastrointestinal diseases
\item[06/D5] Psychiatry
\item[06/D6] Neurology
\end{description}
\item[06/E] Specialized clinical surgery
\begin{description}
\item[06/E1] Heart, thoracic and vascular surgery
\item[06/E2] Plastic and paediatric surgery and urology
\item[06/E3] Neurosurgery and maxillofacial surgery
\end{description}
\item[06/F] Integrated clinical surgery
\begin{description}
\item[06/F1] Odontostomatologic diseases
\item[06/F2] Eye diseases
\item[06/F3] Otorhinolaryngology and audiology
\item[06/F4] Musculoskeletal diseases and physical and rehabilitation medicine
\end{description}
\item[06/G] Paediatrics
\begin{description}
\item[06/G1] Paediatrics and child neuropsychiatry
\end{description}
\item[06/H] Gynaecology
\begin{description}
\item[06/H1] Obstetrics and gynecology
\end{description}
\item[06/I] Radiology
\begin{description}
\item[06/I1] Diagnostic imaging, radiotherapy and neuroradiology
\end{description}
\item[06/L] Anaesthesiology
\begin{description}
\item[06/L1] Anaesthesiology
\end{description}
\item[06/M] Public health
\begin{description}
\item[06/M1] Hygiene, public health, nursing and medical statistics
\item[06/M2] Forensic and occupational medicine
\end{description}
\item[06/N] Applied medical technologies
\begin{description}
\item[06/N1] Applied medical technologies
\end{description}
\end{description}
\item[07] Agricultural and veterinary sciences
\begin{description}
\item[07/A] Agricultural economics and appraisal
\begin{description}
\item[07/A1] Agricultural economics and appraisal
\end{description}
\item[07/B] Agricultural and forest systems
\begin{description}
\item[07/B1] Agronomy and field, vegetable, ornamental cropping systems
\item[07/B2] Arboriculture and forest systems
\end{description}
\item[07/C] Agricultural, forest and biosytems engineering
\begin{description}
\item[07/C1] Agricultural, forest and biosystems engineering
\end{description}
\item[07/D] Plant pathology and entomology
\begin{description}
\item[07/D1] Plant pathology and entomology
\end{description}
\item[07/E] Agricultural chemistry and agricultural genetics
\begin{description}
\item[07/E1] Agricultural chemistry, agricultural genetics and pedology
\end{description}
\item[07/F] Food technology and agricultural microbiology
\begin{description}
\item[07/F1] Food science and technology
\item[07/F2] Agricultural microbiology
\end{description}
\item[07/G] Animal science and technology
\begin{description}
\item[07/G1] Animal science and technology
\end{description}
\item[07/H] Veterinary medicine
\begin{description}
\item[07/H1] Veterinary anatomy and physiology
\item[07/H2] Veterinary pathology and inspection of foods of animal origin
\item[07/H3] Infectious and parasitic animal diseases
\item[07/H4] Clinical veterinary medicine and pharmacology
\item[07/H5] Clinical veterinary surgery and obstetrics
\end{description}
\end{description}
\item[08] Civil engineering and architecture
\begin{description}
\item[08/A] Landscape and infrastructural engineering
\begin{description}
\item[08/A1] Hydraulics, hydrology, hydraulic and marine constructions
\item[08/A2] Sanitary and environmental engineering, hydrocarbons and underground fluids, safety and protection engineering
\item[08/A3] Infrastructural and transportation engineering, real estate appraisal and investment valuation
\item[08/A4] Geomatics
\end{description}
\item[08/B] Structural and geotechnical engineering
\begin{description}
\item[08/B1] Geotechnics
\item[08/B2] Structural mechanics
\item[08/B3] Structural engineering
\end{description}
\item[08/C] Design and technological planning of architecture
\begin{description}
\item[08/C1] Design and technological planning of architecture
\end{description}
\item[08/D] Architectural design
\begin{description}
\item[08/D1] Architectural design
\end{description}
\item[08/E] Drawing, architectural restoration and history
\begin{description}
\item[08/E1] Drawing
\item[08/E2] Architectural restoration and history
\end{description}
\item[08/F] Urban and landscape planning and design
\begin{description}
\item[08/F1] Urban and landscape planning and design
\end{description}
\end{description}
\item[09] Industrial and information engineering
\begin{description}
\item[09/A] Mechanical and aerospace engineering and naval architecture
\begin{description}
\item[09/A1] Aeronautical and aerospace engineering and naval architecture
\item[09/A2] Applied mechanics
\item[09/A3] Industrial design, machine construction and metallurgy
\end{description}
\item[09/B] Manufacturing, industrial and managenent engineering
\begin{description}
\item[09/B1] Manufacturing technology and systems
\item[09/B2] Industrial mechanical plants
\item[09/B3] Business and management engineering
\end{description}
\item[09/C] Energy, thermomechanical and nuclear engineering
\begin{description}
\item[09/C1] Fluid machinery, energy systems and power generation
\item[09/C2] Technical physics and nuclear engineering
\end{description}
\item[09/D] Chemical and materials engineering
\begin{description}
\item[09/D1] Materials science and technology
\item[09/D2] Systems, methods and technologies of chemical and process engineering
\item[09/D3] Chemical plants and technologies
\end{description}
\item[09/E] Electrical and electronic engineering and measurements
\begin{description}
\item[09/E1] Electrical technology
\item[09/E2] Electrical energy engineering
\item[09/E3] Electronics
\item[09/E4] Measurements
\end{description}
\item[09/F] Telecommunications engineering and electromagnetic fields
\begin{description}
\item[09/F1] Electromagnetic fields
\item[09/F2] Telecommunications
\end{description}
\item[09/G] Systems engineering and bioengineering
\begin{description}
\item[09/G1] Systems and control engineering
\item[09/G2] Bioengineering
\end{description}
\item[09/H] Computer engineering
\begin{description}
\item[09/H1] Information processing systems
\end{description}
\end{description}
\item[10] Antiquities, philology, literary studies, art history
\begin{description}
\item[10/A] Archaeological sciences
\begin{description}
\item[10/A1] Archaeology
\end{description}
\item[10/B] Art history
\begin{description}
\item[10/B1] Art history
\end{description}
\item[10/C] Cinema, music, performing arts, television and media studies
\begin{description}
\item[10/C1] Cinema, music, performing arts, television and media studies
\end{description}
\item[10/D] Sciences of antiquity
\begin{description}
\item[10/D1] Ancient history
\item[10/D2] Greek language and literature
\item[10/D3] Latin language and literature
\item[10/D4] Classical and late antique philology
\end{description}
\item[10/E] Medieval latin and romance philologies and literatures
\begin{description}
\item[10/E1] Medieval latin and romance philologies and literatures
\end{description}
\item[10/F] Italian studies and comparative literatures
\begin{description}
\item[10/F1] Italian literature, literary criticism and comparative literature
\item[10/F2] Contemporary Italian literature
\item[10/F3] Italian linguistics and philology
\end{description}
\item[10/G] Glottology and linguistics
\begin{description}
\item[10/G1] Glottology and linguistics
\end{description}
\item[10/H] French studies
\begin{description}
\item[10/H1] French language, literature and culture
\end{description}
\item[10/I] Spanish and Hispanic studies
\begin{description}
\item[10/I1] Spanish and Hispanic languages, literatures and cultures
\end{description}
\item[10/L] English and Anglo-American studies
\begin{description}
\item[10/L1] English and Anglo-American languages, literatures and cultures
\end{description}
\item[10/M] Germanic and Slavic languages, literatures and cultures
\begin{description}
\item[10/M1] Germanic languages, literatures and cultures
\item[10/M2] Slavic studies
\end{description}
\item[10/N] Eastern cultures
\begin{description}
\item[10/N1] Ancient Near Eastern, Middle Eastern and African cultures
\item[10/N3] Central and East Asian cultures
\end{description}
\end{description}
\item[11] History, philosophy, pedagogy and psychology
\begin{description}
\item[11/A] History
\begin{description}
\item[11/A1] Medieval history
\item[11/A2] Modern history
\item[11/A3] Contemporary history
\item[11/A4] Science of books and documents, history of religions
\item[11/A5] Demography, ethnography and anthropology
\end{description}
\item[11/B] Geography
\begin{description}
\item[11/B1] Geography
\end{description}
\item[11/C] Philosophy
\begin{description}
\item[11/C1] Theoretical philosophy
\item[11/C2] Logic, history and philosophy of science
\item[11/C3] Moral philosophy
\item[11/C4] Aesthetics and philosophy of languages
\item[11/C5] History of philosophy
\end{description}
\item[11/D] Educational theories
\begin{description}
\item[11/D1] Educational theories and history of educational theories
\item[11/D2] Methodologies of teaching, special education and educational research
\end{description}
\item[11/E] Psychology
\begin{description}
\item[11/E1] General psychology, psychobiology and psychometrics
\item[11/E2] Developmental and educational psychology
\item[11/E3] Social psychology and work and organizational psychology
\item[11/E4] Clinical and dynamic psychology
\end{description}
\end{description}
\item[12] Law studies
\begin{description}
\item[12/A] Private law
\begin{description}
\item[12/A1] Private law
\end{description}
\item[12/B] Business, navigation and air law and labour law
\begin{description}
\item[12/B1] Business, navigation and air law
\item[12/B2] Labour law
\end{description}
\item[12/C] Constitutional and ecclesiastical law
\begin{description}
\item[12/C1] Constitutional law
\item[12/C2] Ecclesiastical law and canon law
\end{description}
\item[12/D] Administrative and tax law
\begin{description}
\item[12/D1] Administrative law
\item[12/D2] Tax law
\end{description}
\item[12/E] International and European Union law, comparative, economics and markets law
\begin{description}
\item[12/E1] International and European Union law
\item[12/E2] Comparative law
\item[12/E3] Economics, financial and agri-food markets law and regulation
\end{description}
\item[12/F] Civil procedural law
\begin{description}
\item[12/F1] Civil procedural law
\end{description}
\item[12/G] Criminal law and criminal procedure
\begin{description}
\item[12/G1] Criminal law
\item[12/G2] Criminal procedure
\end{description}
\item[12/H] Roman law, history of medieval and modern law and philosophy of law
\begin{description}
\item[12/H1] Roman and ancient law
\item[12/H2] History of medieval and modern law
\item[12/H3] Philosophy of law
\end{description}
\end{description}
\item[13] Economics and statistics
\begin{description}
\item[13/A] Economics
\begin{description}
\item[13/A1] Economics
\item[13/A2] Economic policy
\item[13/A3] Public economics
\item[13/A4] Applied economics
\item[13/A5] Econometrics
\end{description}
\item[13/B] Business administration and Management
\begin{description}
\item[13/B1] Business administration and Management
\item[13/B2] Management
\item[13/B3] Organization studies
\item[13/B4] Financial Markets and Institutions
\item[13/B5] Commodity science
\end{description}
\item[13/C] Economic history
\begin{description}
\item[13/C1] Economic history
\end{description}
\item[13/D] Statistics and mathematical methods for decisions
\begin{description}
\item[13/D1] Statistics
\item[13/D2] Economic statistics
\item[13/D3] Demography and social statistics
\item[13/D4] Mathematical methods of economics, finance and actuarial sciences
\end{description}
\end{description}
\item[14] Political and social sciences
\begin{description}
\item[14/A] Political theory
\begin{description}
\item[14/A1] Political philosophy
\item[14/A2] Political science
\end{description}
\item[14/B] Political history
\begin{description}
\item[14/B1] History of political thought and institutions
\item[14/B2] History of international relations and of non-European societies and institutions
\end{description}
\item[14/C] Sociology
\begin{description}
\item[14/C1] General and political sociology, sociology of law
\item[14/C2] Sociology of culture and communication
\end{description}
\item[14/D] Applied sociology
\begin{description}
\item[14/D1] Sociology of economy and labour, sociology of land and environment\end{description}
\end{description}
\end{description}

\bibliographystyle{ACM-Reference-Format}
\bibliography{wse}

\end{document}